\documentclass[12pt]{article}
\usepackage{amsmath}
\usepackage{amssymb}
\usepackage{amsfonts}
\usepackage{latexsym}
\usepackage{color}
\usepackage{graphicx}

\catcode `\@=11 \@addtoreset{equation}{section}

\catcode `\@=12

\newcommand{\be}{\begin{equation}}
\newcommand{\en}{\end{equation}}
\newcommand{\bea}{\begin{eqnarray}}
\newcommand{\ena}{\end{eqnarray}}
\newcommand{\beano}{\begin{eqnarray*}}
\newcommand{\enano}{\end{eqnarray*}}
\newcommand{\bee}{\begin{enumerate}}
\newcommand{\ene}{\end{enumerate}}

\newcommand{\Hil}{{\cal H}}

\newcommand{\J}{{\cal J}}

\newcommand{\Id}{1\!\!1}
\newcommand{\F}{{\cal F}}
\newcommand{\Lc}{{\cal L}}

\newcommand{\G}{{\cal G}}
\newcommand{\E}{{\cal E}}

\newtheorem{thm}{Theorem}

\newtheorem{prop}[thm]{Proposition}

\begin{document}

\thispagestyle{empty}

\vspace*{1cm}

\begin{center}
{\Large \bf Intertwining operators between different Hilbert spaces: connection with frames}   \vspace{2cm}\\

{\large F. Bagarello}\\
  Dipartimento di Metodi e Modelli Matematici,
Facolt\`a di Ingegneria, Universit\`a di Palermo, I-90128  Palermo, Italy\\
e-mail: bagarell@unipa.it
\end{center}


\vspace*{2cm}

\begin{abstract}
\noindent In this paper we generalize a strategy recently proposed by the author concerning intertwining operators. In particular we discuss the possibility of extending our previous results in such a way to  construct (almost) isospectral self-adjoint operators living in different Hilbert spaces. Many examples are discussed in details. Many of them arise from the theory of frames in Hilbert spaces, others from the so-called g-frames.

\end{abstract}

\vspace{2cm}

\vfill

\newpage

\section{Introduction}

In two recent papers, \cite{bag1,bag2}, we have proposed a new technique which produces, given few ingredients, a  hamiltonian $h_2$ which has (almost) the same spectrum of a given hamiltonian $h_1$ and whose respective eigenstates are related  by a given intertwining operator (IO). More precisely, calling $\sigma(h_j)$, $j=1,2$, the set of eigenvalues of $h_j$, we find that $\sigma(h_2)\subseteq\sigma(h_1)$. These results extend what was discussed in the previous literature on this subject, \cite{intop}, and have the advantage of being a constructive procedure: while in \cite{intop} the existence of $h_1, h_2$ and of an operator $x$ satisfying the  intertwining condition $h_1x=xh_2$ is assumed, in \cite{bag1,bag2} we explicitly construct $h_2$ from $h_1$ and $x$ in such a way that $h_2$ satisfies a {\em weak form} of  $h_1x=xh_2$. Moreover, as mentioned above, $\sigma(h_2)\subseteq\sigma(h_1)$ and the eigenvectors are related in a standard way, see \cite{bag1,bag2}. It is well known that this procedure is strongly related to, and actually extends, the supersymmetric quantum mechanics widely discussed in the past years, see \cite{CKS} and \cite{jun} for interesting  reviews.

In \cite{bag1,bag2} we have considered the  relation between our IO technique and vector Gazeau-Klauder like coherent states. More in details, we have shown how a certain kind of coherent states can be defined by isospectral hamiltonians and, vice-versa, which kind of isospectral hamiltonians  arise from certain generalized coherent states. In the cited paper, and in the standard literature on the subject, the two partner hamiltonian operators $h_1$ and $h_2$ acts on the same Hilbert space. Here  we remove this constraint showing that our procedure can be safely extended to certain pairs of operators acting on different Hilbert spaces. This is the main contain of Section II. In Section III we discuss the relation between the theory of frames in Hilbert spaces, \cite{chri,dau},  and the results of Section II. In particular we use tight frames to construct examples of partner hamiltonians in different Hilbert spaces. Section IV contains more examples arising from a generalized version of frames, the so called g-frames, \cite{sun,najati}. In order to keep the paper self-contained, few relevant aspects of frames and g-frames are also discussed in Sections III and IV. Section V contains our conclusions.

\section{IOs between different Hilbert spaces}

 We begin this section briefly sketching what we have done in \cite{bag1}. Let $h_1$ be a self-adjoint hamiltonian on the Hilbert space $\Hil$, $h_1=h_1^\dagger$, whose normalized eigenvectors, $\hat\varphi_n^{(1)}$, satisfy the following equation: $h_1\hat\varphi_n^{(1)}=\epsilon_n\hat\varphi_n^{(1)}$, $n\in\Bbb{N}$. Suppose that there exists an operator $x_1$ on $\Hil$ with the following properties: $
[x_1x_1^\dagger,h_1]=0,
$
and  $N:=x_1^\dagger\,x_1$ is invertible.  Then, calling
$h_2:=N^{-1}\left(x_1^\dagger\,h_1\,x_1\right)$  and $\varphi_n^{(2)}=x_1^\dagger\hat\varphi_n^{(1)},
$
we find that
$h_2=h_2^\dagger$,
$ x_1^\dagger\left(x_1\,h_2-h_1\,x_1\right)=0$ and, if
$\varphi_n^{(2)}\neq 0$, then $h_2\varphi_n^{(2)}=\epsilon_n\varphi_n^{(2)}.$
Notice that, contrarily to $\hat\varphi_n^{(1)}$, in general $\varphi_n^{(2)}$ is not normalized.
Of course, all the operators appearing here act on $\Hil$ and both $\hat\varphi_n^{(1)}$ and $\varphi_n^{(2)}$ belong to $\Hil$ itself for all $n\in\Bbb{N}$. However we have several examples both in physical and in mathematical literature of operators $X$ acting between different Hilbert spaces and such that $X^\dagger X=A\Id$, $A\in\Bbb{R}$. Just to mention one, this is exactly what happens if $X$ is the {\em analysis} operator of a tight frame, \cite{chri}. In this case $X$ maps a certain Hilbert space $\Hil$ into $l^2(\Bbb{N})$.  Other examples are discussed, for instance, in \cite{alibag}, in connection with the physics of Landau levels. Still more examples arise from g-frames. These examples motivate our analysis, which is just a natural extension of the above results to the slightly different situation we want to consider here.

Let $h_1$ be a self-adjoint hamiltonian on the Hilbert space $\Hil_1$, $h_1=h_1^\dagger$, whose  eigenvectors, $\varphi_n^{(1)}$, satisfy the following equation: $h_1\varphi_n^{(1)}=\epsilon_n\varphi_n^{(1)}$, $n\in\Bbb{N}$\footnote{In what follows we will not be particularly interested to the normalization of the various eigenstates of $h_1$ and $h_2$.}. Let $\Hil_2$ be a second Hilbert space, in general different from $\Hil_1$, and consider an operator $X:\Hil_2\rightarrow\Hil_1$, whose adjoint $X^\dagger$ maps $\Hil_1$ in $\Hil_2$. Let us further define
\be
N_1:=XX^\dagger,\qquad N_2:=X^\dagger X
\label{24}
\en
It is clear that $N_j$ maps $\Hil_j$ into itself, for $j=1,2$. Suppose now that $X$ is such that $N_2$ is invertible in $\Hil_2$ and
\be
[N_1,h_1]=0.
\label{25}\en
Of course this commutator should be considered in a {\em weak form} if $h_1$ or $N_1$ is unbounded: $<N_1f,h_1g>=<h_1f,N_1g>$, for $f,g$ in the domain of $N_1$ and $h_1$. Defining now
\be
h_2:=N_2^{-1}\left(X^\dagger\,h_1\,X\right),\qquad \varphi_n^{(2)}=X^\dagger\varphi_n^{(1)},
\label{26}\en
 the following conditions are satisfied:
\be \left\{
\begin{array}{ll}
\mbox{$[\alpha]$}\qquad h_2=h_2^\dagger  \\
\mbox{$[\beta]$}\qquad X^\dagger\left(X\,h_2-h_1\,X\right)=0\\
\mbox{$[\gamma]$}\qquad \mbox{if }\varphi_n^{(2)}\neq 0 \mbox{ then } h_2\varphi_n^{(2)}=\epsilon_n\varphi_n^{(2)}.\\
\end{array}
\right. \label{27} \en
The proof of these statements do not differ significantly from that in \cite{bag1} and will not be repeated here. Notice that condition $[\gamma]$ implies that $\epsilon_n$ is, if $\varphi_n^{(2)}\neq 0$, an eigenvalue of both $h_1$ and $h_2$. In the following we will say that $h_1$ and $h_2$ are {\em almost isospectral}, since in general we will get that $\sigma(h_2)\subseteq\sigma(h_1)$. Let us see in some details few consequences of our approach. First of all, we notice that $h_2$ is an operator acting on $\Hil_2$.  Moreover it is possible to check that $[h_2,N_2]=0$\footnote{Again we will neglect all the difficulties arising from the fact that $h_2$ or $N_2$ could be unbounded. So, in a sense, $[h_2,N_2]=0$ is just a formal expression.}. To prove this equality, we first observe that $N_2h_2=X^\dagger\,h_1\,X$. Moreover, using our definitions and the fact that $[N_1,h_1]=0$, we get
$$
h_2N_2=N_2^{-1}\left(X^\dagger\,h_1\,X\right)\left(X^\dagger\,X\right)=N_2^{-1}X^\dagger\,h_1\,N_1\,X
=N_2^{-1}X^\dagger\,N_1\,h_1\,X=\\$$
$$=N_2^{-1}\,N_2\,X^\dagger\,h_1\,X=X^\dagger\,h_1\,X,
$$
which is equal to $N_2h_2$. Hence, if $[h_1,N_1]=0$ and if $h_2:=N_2^{-1}\left(X^\dagger\,h_1\,X\right)$, then $[h_2,N_2]=0$. In other words, $h_1$ and $h_2$ have a kind of symmetric behavior, which is also reflected from the following result:

if $h_2$ is defined as in (\ref{26}) and if ${N_1}^{-1}$ exists, then
\be
h_1:=N_1^{-1}\left(X\,h_2\,X^\dagger\right)
\label{28}\en
Indeed left-multiplying (\ref{26}) for $XN_2$ and using the existence of
${N_1}^{-1}$, together with $[h_2,N_2]=0$, simple algebraic manipulations show that $h_1X={N_1}^{-1}
\left(X\,h_2\,X^\dagger\right)X$ or, in other words, that $\left(h_1-{N_1}^{-1}
\left(X\,h_2\,X^\dagger\right)\right)X=0$. Right-multiplying this equation for $X^\dagger$ and using the invertibility of $N_1=XX^\dagger$ we recover (\ref{28}).

Moreover, under the same conditions it is possible to check that $[\beta]$ holds in the stronger standard form: $Xh_2=h_1X$. Indeed, it is sufficient to left-multiply $[\beta]$ for $X$ and to use the existence of $N_1^{-1}$.

\vspace{2mm}

{\bf Remark:--} The above requirement is strong in the sense that it is required the existence of both $N_1^{-1}$ and $N_2^{-1}$. This hypothesis can be replaced assuming that $h_1$, $h_2$ and $X$ satisfy the usual weak intertwining relation  $X^\dagger\left(X\,h_2-h_1\,X\right)=0$ and that $[h_2,N_2]=0$:  these two assumptions again produce formula (\ref{28}).

\vspace{2mm}

Another {\em inversion result} concerns the relation between the eigenstates of $h_1$ and $h_2$. We have seen in $[\gamma]$, (\ref{27}), that, if $\varphi_n^{(2)}=X^\dagger \varphi_n^{(1)}\neq 0$  then $h_2\varphi_n^{(2)}=\epsilon_n\varphi_n^{(2)}$.
It is also possible to show that $X\varphi_n^{(2)}$ is an eigenstate of $h_1$ with eigenvalue $\epsilon_n$. Hence, if $\epsilon_n$ is not degenerate, $X\varphi_n^{(2)}$ is proportional to $\varphi_n^{(1)}$: $X\varphi_n^{(2)}=\alpha_n\varphi_n^{(1)}$, for some $\alpha_n$. This also implies that $X\,X^\dagger\varphi_n^{(1)}=N_1\varphi_n^{(1)}=\alpha_n\,\varphi_n^{(1)}$. Hence $\varphi_n^{(1)}$ is an eigenstate of both $h_1$ and $N_1$. This is not a big surprise, since these two operators commute and therefore can be simultaneously diagonalized.

The roles of $h_1$ and $h_2$ can be easily exchanged: suppose that $h_1:=N_1^{-1}\left(X\,h_2\,X^\dagger\right)$, $[h_2,N_2]=0$ and let $\varphi_n^{(2)}$ be an eigenstate of $h_2$ with eigenvalue $\epsilon_n$, then $X\varphi_n^{(2)}$ is an eigenstate of $h_1$ with the same eigenvalue. The above discussion can be summarized by the following proposition, which also contains some extra results.

\begin{prop}
Let $\Hil_1$ and $\Hil_2$ be two Hilbert spaces and $X:\Hil_2\rightarrow\Hil_1$. Let us put $N_1:=XX^\dagger$ and $N_2:=X^\dagger X$.

If $h_1$ is a self-adjoint operator on $\Hil_1$ commuting with $N_1$, $[N_1,h_1]=0$, and if $N_2$ admits inverse in $\Hil_2$, then calling $h_2=N_2^{-1}\left(X^\dagger\,h_1\,X\right)$, if $\varphi_n^{(1)}\in\Hil_1$ is such that $h_1\varphi_n^{(1)}=\epsilon_n\varphi_n^{(1)}$ and  $X^\dagger\varphi_n^{(1)}$ is not zero, we have $h_2\left(X^\dagger\varphi_n^{(1)}\right)=\epsilon_n\left(X^\dagger\varphi_n^{(1)}\right)$. Moreover we have $h_2=h_2^\dagger$, $X^\dagger(Xh_2-h_1X)=0$ and $[h_2,N_2]=0$. Furthermore, if $\epsilon_n$ is non degenerate, then $\varphi_n^{(1)}$  and $X^\dagger\varphi_n^{(1)}$ are eigenstates of $N_1$ and $N_2$ respectively with the same eigenvalue.

If, on the other way,  $N_1$ admits inverse in $\Hil_1$, taken a self-adjoint operator $h_2$ on $\Hil_2$ such that $[N_2,h_2]=0$  we can define the following operator on $\Hil_1$: $h_1=N_1^{-1}\left(X\,h_2\,X^\dagger\right)$. Now, if $\varphi_n^{(2)}\in\Hil_2$ is such that $h_2\varphi_n^{(2)}=\epsilon_n\varphi_n^{(2)}$ and $X\varphi_n^{(2)}$ is not zero, we have $h_1\left(X\varphi_n^{(2)}\right)=\epsilon_n\left(X\varphi_n^{(2)}\right)$. Moreover it turns out that $h_1=h_1^\dagger$, $X(X^\dagger\,h_1-h_2X^\dagger)=0$ and $[h_1,N_1]=0$. Furthermore, if $\epsilon_n$ is non degenerate, then $\varphi_n^{(2)}$  and $X\,\varphi_n^{(2)}$ are eigenstates of $N_2$ and $N_1$ respectively with the same eigenvalue.

\end{prop}

 We give here a simple example of this Proposition, while in the next section other more interesting situations will be considered.

\vspace{2mm}

{\bf A first introductory example:--}
Let $h_1=h_1^\dagger$ be an hamiltonian operator on $\Hil_1$ and let $V$ be an unitary map between $\Hil_1$ and $\Hil_2$. Let us define a second self-adjoint operator $h_2$ on $\Hil_2$ as $h_2=Vh_1V^{-1}$. It is well known that $\sigma(h_1)=\sigma(h_2)$ and that, if $\varphi_n^{(1)}$ satisfies the eigenvalue equation $h_1\varphi_n^{(1)}=\epsilon_n\varphi_n^{(1)}$, then $\varphi_n^{(2)}:=V\varphi_n^{(1)}$ satisfies the analogous equation  $h_2\varphi_n^{(2)}=\epsilon_n\varphi_n^{(2)}$. This can be seen as a special case of our Proposition. Indeed, if we take $X=V^\dagger$, we get $N_1:=V^\dagger V=\Id_{\Hil_1}$ and $N_2:=VV^\dagger =\Id_{\Hil_2}$. Then, for instance, $[h_1,N_1]=0$ and $h_2=Vh_1V^\dagger$ or, equivalently, $h_2V=Vh_1$. Hence $V$ is a standard IO. Moreover $[h_2,N_2]=0$.  We refer to \cite{alibag} for a non trivial example of this situation arising from Landau levels and related to a concrete realization of the Tomita-Takesaki modular structure. In particular the unitary map considered in \cite{alibag} acts between the Hilbert space of the trace-class operators on $\Lc^2(\Bbb{R})$, $B_2(\Lc^2(\Bbb{R}))$, and $\Lc^2(\Bbb{R}^2)$.

\vspace{2mm}

{\bf Remark:--} Of course, a class of examples of our construction comes from the differential IOs considered in the existing literature on supersymmetric quantum mechanics for which, however, $\Hil_1=\Hil_2$. These will not be considered here. We will  add few comments in Section V.

\section{Examples from tight frames}

In this section we will show how the general theory of frames can be used to construct various examples which fit the assumptions of our settings. For that we first recall few important facts and definitions on frames, see \cite{chri,dau}.

Let $\Hil$ be a separable Hilbert
space (which could also be finite-dimensional) and $\F\equiv \{\varphi_n\in\Hil,\, n\in \Bbb{N}\}$, be a
set of vectors of $\Hil$. We say that $\F$ is an (A,B)-{frame} of $\Hil$ if there
exist two positive constants, called frame bounds, $0<A\leq B<\infty$, such that the
inequalities\be A\|f\|^2 \leq \sum_{n\in \Bbb{N}}|<\varphi_n,f>|^2 \leq B\|f\|^2
\label{31}
\en
hold for any $f\in \Hil$. In the case of finite dimensional $\Hil$ we just replace $\Bbb{N}$ with a finite subset of the natural numbers.

To any such set $\F$ can be associated a bounded operator $F:\Hil \rightarrow {
l}^2(\Bbb{N})=\{\{c_n\in\Bbb{C}\}_{n\in \Bbb{N}}\, : \: \sum_{n\in \Bbb{N}}|c_n|^2\}<\infty$, called the {\em analysis operator},  defined by the formula
\be
\forall f\in \Hil \hspace{1cm}  (Ff)_j = <\varphi_j, f> \qquad\Longrightarrow\qquad Ff=\left\{<\varphi_j, f>\right\}_{j\in \Bbb{N}}
\label{32}
\en
Due to equation (\ref{31}) we see that $Ff\in l^2(\Bbb{N})$ and $\|F\| \leq \sqrt{B}$. The
adjoint of the operator $F$, the so-called {\em synthesis operator} $F^\dagger$,  maps ${ l}^2(\Bbb{N})$ into $\Hil$ and satisfies
 \be
\forall \{c\} \in { l}^2(\Bbb{N}) \hspace{1cm}  F^\dagger c \equiv \sum_{i\in \Bbb{N}}c_i \varphi_i
\label{33}
\en

By means of these operators condition (\ref{31}) can be rewritten in the following
equivalent way: $\F$ is an (A,B)-{ frame} of $\Hil$ if there exist
two positive constants, $0<A\leq B<\infty$, such that the inequalities
\be
A\Id \leq F^\dagger F \leq B \Id
\label{34}
\en
hold in the sense of the operators, \cite{rs}. We have used $\Id$ to identify the
identity operator on $\Hil$.

Condition (\ref{34}) implies that the {\em frame operator} $F^\dagger F$, which maps $\Hil$ into itself, can
be inverted and that its inverse, $(F^\dagger F)^{-1}$, is still bounded in $\Hil$. In other
terms, we have that both $F^\dagger F$ and $(F^\dagger F)^{-1}$ belong to $B(\Hil)$, the set of all bounded operators on $\Hil$.

Following the literature, see \cite{dau} for instance, one may introduce the {\em dual frame} of
$\F$, $\tilde \F$, as the set of vectors $\tilde \varphi_i$ defined by
$\tilde \varphi_i \equiv (F^\dagger F)^{-1} \varphi_i$, $\forall i\in \Bbb{N}$.
$\tilde \F$ is  a $(\frac{1}{B},\frac{1}{A})$-frame, see \cite{dau}, where it is also proved that  any vector of the Hilbert space can be expanded as
linear combinations of the vectors of the set $\F$ or of the set $\tilde \F$. We have the
following reconstruction formulas:
\be
f=\sum_{i\in \Bbb{N}}<\varphi_i,f> \tilde \varphi_i = \sum_{i\in \Bbb{N}}<\tilde \varphi_i,f> \varphi_i
\label{36}
\en
for all $f\in \Hil$. More details can be found in \cite{chri,dau}.

As it is clear from equation (\ref{36}), a crucial role in the reconstruction procedure is
the knowledge of the set $\tilde \F$. In order to obtain the explicit expression for $\tilde
\varphi_i$, we first have to know how the operator $(F^*F)^{-1}$ acts on the vectors of
$\Hil$. This is, in general, a difficult problem to solve. Only in a single situation we can
give an easy answer, namely when our frame is {\em tight}. This means that the frame bounds
$A$ and $B$ coincide, $A=B$, so that equation (\ref{34})
reduces to
\be
 F^\dagger F = A \Id,
\label{37}
\en
which implies also that $(F^\dagger F)^{-1}=\frac{1}{A} \Id$. Therefore $\tilde \varphi_i=
\frac{1}{A} \varphi_i$, for all $i\in \Bbb{N}$. In this case the reconstruction formulas above
coincide and they look like
$
f=\frac{1}{A} \sum_{i\in \Bbb{N}}<\varphi_i,f> \varphi_i,
$
for all $f\in \Hil$. In particular, moreover, if $A=1$ the set $\F$ is called a Parceval tight frame and if all the vectors $\varphi_i$
are normalized, it follows that this frame forms an orthonormal basis of the Hilbert
space, see \cite{dau}. Of course, also the vice-versa holds true: if $\F$ is an orthonormal
set in $\Hil$, then $\F$ is a $(1,1)$-frame of normalized vectors. This is an obvious
consequence of the Parceval equality $\sum_n|<\varphi_n,f>|^2 =\|f\|^2$, which holds for all
$f\in \Hil$.

\vspace{2mm}

What is relevant for us is the fact that $F$  maps a given Hilbert space $\Hil$ into a different one $l^2(\Bbb{N})$ and is such that, if the frame is tight, $F^\dagger F$ is just a multiple of the identity operator on $\Hil$ and therefore commutes with any possible operator acting on $\Hil$. This means that if we take the intertwining operator $X=F^\dagger$, then $\Hil_1\equiv\Hil$ and $\Hil_2=l^2(\Bbb{N})$ and, for all possible choices of $h_1$ we have automatically  $[h_1,N_1]=0$. We only have to check whether $N_2=FF^\dagger$ admits inverse or not. Since $N_2\geq0$, it is enough to check if $\ker\{F^\dagger\}$ contains only the zero vector. However, how we will show in the examples below, $N_2$ is usually not an invertible operator. Hence a different approach should be considered, if possible. For this reason we now give two different possibilities, both related to tight frames, which are in a sense complementary: if one cannot be used then we can surely use the other  in order to produce almost isospectral hamiltonians living in different Hilbert spaces.

\vspace{3mm}

{\bf Option I:} this is when $FF^\dagger$ is an invertible operator in $l^2(\Bbb{N})$. If this is so then we put
\be
\Hil_1\equiv\Hil,\qquad \Hil_2\equiv l^2(\Bbb{N}), \qquad X=F^\dagger,\quad\mbox{and}\quad X^\dagger=F
\label{38}\en
Hence $X:l^2(\Bbb{N})\rightarrow \Hil$ and $X^\dagger:\Hil\rightarrow l^2(\Bbb{N})$, and $N_1:=XX^\dagger=F^\dagger F =A\Id_\Hil$ (since $\F$ is a tight frame), $N_2:=X^\dagger X=FF^\dagger:l^2(\Bbb{N})\rightarrow l^2(\Bbb{N})$. Here we are using $\Id_\Hil$ to indicate the identity operator in $\Hil$. Analogously we will use $\Id_{l^2}$ for the identity operator in $l^2(\Bbb{N})$.

With these definitions we deduce that, as already mentioned, any self-adjoint operator $h_1$ on $\Hil$ will commute with $N_1$ so that condition (\ref{25}) is satisfied. Also, since we are here assuming that $N_2=FF^\dagger$ is an invertible operator in $l^2(\Bbb{N})$, also the second assumption of our construction is verified. Hence we can introduce a second  operator $h_2$ on $l^2(\Bbb{N})$ as in (\ref{26}), $h_2:=N_2^{-1}\left(X^\dagger\,h_1\,X\right)$.
We know that $h_2$ is self-adjoint, $F\left(F^\dagger\,h_2-h_1\,F^\dagger\right)=0$ and that $[h_2,N_2]=0$. Moreover, calling $\varphi_n^{(1)}$ the eigenstate of $h_1$ with eigenvalue $\epsilon_n$,  if $F\,\varphi_n^{(1)}$ is not zero, then  $h_2(F\varphi_n^{(1)})=\epsilon_n(F\varphi_n^{(1)})$. We can further check that $\varphi_n^{(1)}$ is also an eigenstate of $N_1$ with eigenvalue $A$ (and this is trivial), and $F\varphi_n^{(1)}$ is an eigenstate of $N_2$ with the same eigenvalue.

\vspace{3mm}

As we have already mentioned, in many examples the operator $N_2$  turns out not to be invertible. Hence Option I cannot be used. However,  frames can still be used in a slightly different way. In fact, because of the definition of a general $(A,B)-$frame, see (\ref{34}), $F^\dagger F$ is surely invertible in $\Hil$, even if $\F$ is not tight. This suggests the introduction of the following alternative possibility.

{\bf Option II:} this is when $FF^\dagger$ is not an invertible operator in $l^2(\Bbb{N})$. In this case  we put
\be
\Hil_1\equiv l^2(\Bbb{N}),\qquad \Hil_2\equiv\Hil , \qquad X=F,\quad\mbox{and}\quad X^\dagger=F^\dagger
\label{39}\en
Hence $X: \Hil\rightarrow l^2(\Bbb{N})$ and $X^\dagger:l^2(\Bbb{N})\rightarrow \Hil$, and $\mathfrak{N}_1:=XX^\dagger=F F^\dagger:l^2(\Bbb{N})\rightarrow l^2(\Bbb{N})$, $\mathfrak{N}_2:=X^\dagger X=F^\dagger F$. Of course, if we take as before $\F$ as a tight frame, $\mathfrak{N}_2=A\Id_\Hil$. It is clear then that, with these choices, $\mathfrak{N}_2$ admits inverse. The difficulty is here in finding the self adjoint operator $h_1$ on $l^2(\Bbb{N})$ which commutes with $\mathfrak{N}_1=F F^\dagger$. However, if such an operator can be found, then we define as usual $h_2:=\mathfrak{N}_2^{-1}\left(X^\dagger\,h_1\,X\right)=\frac{1}{A}\,F^\dagger\,h_1\,F$, which is an operator on $\Hil$. Also, if $\varphi_n^{(1)}$ the eigenstate of $h_1$ with eigenvalue $\epsilon_n$, and if $F^\dagger\,\varphi_n^{(1)}$ is not zero, then it is an eigenstate of $h_2$ with the same eigenvalue: $h_2(F^\dagger\varphi_n^{(1)})=\epsilon_n(F^\dagger\varphi_n^{(1)})$. As before,   $\varphi_n^{(2)}=F^\dagger\varphi_n^{(1)}$ is also trivially an eigenstate of $\mathfrak{N}_2$ with eigenvalue $A$, while $F\varphi_n^{(2)}$ is an eigenstate of $\mathfrak{N}_1$ with the same eigenvalue.

\subsection{A  finite dimensional example}

Let $\F=\{\chi_j\in\Bbb{C}^3,\,j=1,2,3,4,5\}$ be the following set of three-dimensional vectors:
$$\chi_1=\frac{1}{\sqrt{3}}\left(
                                                                              \begin{array}{c}
                                                                                0 \\
                                                                                1 \\
                                                                                \sqrt{2} \\
                                                                              \end{array}
                                                                            \right),\, \chi_2=\frac{1}{\sqrt{3}}\left(
                                                                              \begin{array}{c}
                                                                                0 \\
                                                                                -1 \\
                                                                                \sqrt{2} \\
                                                                              \end{array}
                                                                            \right),\, \chi_3=\left(
                                                                              \begin{array}{c}
                                                                                0 \\
                                                                                1 \\
                                                                                0\\
                                                                              \end{array}
                                                                            \right),$$
                                                                            $$ \chi_4=\frac{1}{\sqrt{6}}\left(
                                                                              \begin{array}{c}
                                                                                \sqrt{5} \\
                                                                                0 \\
                                                                                1 \\
                                                                              \end{array}
                                                                            \right),\,\chi_5=\frac{1}{\sqrt{3}}\left(
                                                                              \begin{array}{c}
                                                                                -\sqrt{5} \\
                                                                                0 \\
                                                                                1 \\
                                                                              \end{array}
                                                                            \right).
$$
It is known, \cite{chri}, that $\F$ is a tight frame in $\Bbb{C}^3$ with $A=\frac{5}{3}$. We find the following matricial forms for $F:\Bbb{C}^3\rightarrow\Bbb{C}^5$ and its adjoint  $F^\dagger:\Bbb{C}^5\rightarrow\Bbb{C}^3$:
$$
F=\left(
    \begin{array}{ccc}
      0 & \frac{1}{\sqrt{3}} & \sqrt{\frac{2}{3}} \\
      0 & -\frac{1}{\sqrt{3}} & \sqrt{\frac{2}{3}} \\
      0 & 1 & 0 \\
      \sqrt{\frac{5}{6}} & 0 & \frac{1}{\sqrt{6}} \\
      -\sqrt{\frac{5}{6}} & 0 & \frac{1}{\sqrt{6}} \\
    \end{array}
  \right),\qquad F^\dagger=\left(
                             \begin{array}{ccccc}
                               0 & 0 & 0 & \sqrt{\frac{5}{6}} & -\sqrt{\frac{5}{6}} \\
                               \frac{1}{\sqrt{3}} & -\frac{1}{\sqrt{3}} & 1 & 0 & 0 \\
                               \sqrt{\frac{2}{3}} & \sqrt{\frac{2}{3}} & 0 & \frac{1}{\sqrt{6}} & \frac{1}{\sqrt{6}} \\
                             \end{array}
                           \right).
$$
Hence
$$
F^\dagger F=\frac{5}{3}\Id_{\Bbb{C}^3}, \qquad FF^\dagger=\left(
                                                            \begin{array}{ccccc}
                                                              1 & \frac{1}{3} & \frac{1}{\sqrt{3}} & \frac{1}{3} & \frac{1}{3} \\
                                                              \frac{1}{3} & 1 & -\frac{1}{\sqrt{3}} & \frac{1}{3} & \frac{1}{3} \\
                                                              \frac{1}{\sqrt{3}} & -\frac{1}{\sqrt{3}} & 1 & 0 & 0 \\
                                                              \frac{1}{3} & \frac{1}{3} & 0 & 1 & -\frac{2}{3} \\
                                                              \frac{1}{3} & \frac{1}{3} & 0 & -\frac{2}{3} & 1 \\
                                                            \end{array}
                                                          \right)
$$
Because of the redundancy of $\F$ in $\Bbb{C}^3$ $FF^\dagger$ does not admit inverse. Indeed we have $\sigma(FF^\dagger)=\{\frac{5}{3},\frac{5}{3},\frac{5}{3},0,0\}$ which contains $\sigma(F^\dagger\,F)=\{\frac{5}{3},\frac{5}{3},\frac{5}{3}\}$ as a proper subset. Hence we are  forced to use Option II above. We begin defining $\mathfrak{N}_1:=F F^\dagger$ and $\mathfrak{N}_2=F^\dagger F=\frac{5}{3}\Id_{\Bbb{C}^3}$. A $5\times5$ self-adjoint matrix commuting with $\mathfrak{N}_1$ is the following:
$$
h_1=\left(
      \begin{array}{ccccc}
        \frac{1}{15}(43+6\sqrt{3}) & -\frac{14}{15} & \frac{2}{5}(-1+\sqrt{3}) & -\frac{2}{5}(-1+\sqrt{3}) & \frac{1}{15}(1-6\sqrt{3}) \\
        -\frac{14}{15} & \frac{1}{15}(43-6\sqrt{3}) & -\frac{2}{5}(1+\sqrt{3}) & \frac{2}{5}(1+\sqrt{3}) & \frac{1}{15}(1+6\sqrt{3}) \\
        \frac{2}{5}(-1+\sqrt{3}) & -\frac{2}{5}(1+\sqrt{3}) & \frac{21}{5} & \frac{4}{5} & \frac{4}{5} \\
        -\frac{2}{5}(-1+\sqrt{3}) & \frac{2}{5}(1+\sqrt{3}) & \frac{4}{5} & \frac{11}{5} & -\frac{4}{5} \\
        \frac{1}{15}(1-6\sqrt{3}) & \frac{1}{15}(1+6\sqrt{3}) & \frac{4}{5} & -\frac{4}{5} & \frac{28}{15} \\
      \end{array}
    \right)
$$
which we now take as our  hamiltonian $h_1$. Hence $\sigma(h_1):=\{5,2+\sqrt{5},3,2,2-\sqrt{2}\}$ and the related eigenstates are partly given below:
$$\varphi_1^{(1)}=\left(
                                                                              \begin{array}{c}
                                                                                \frac{9+\sqrt{3}}{3+9\sqrt{3}} \\
                                                                                -\frac{9+\sqrt{3}}{3+9\sqrt{3}} \\
                                                                                1 \\
                                                                                0\\
                                                                                0\\
                                                                              \end{array}
                                                                            \right),\qquad \varphi_3^{(1)}=-\frac{1}{2}\left(
                                                                              \begin{array}{c}
                                                                                1 \\
                                                                                1 \\
                                                                                0 \\
                                                                                3\\
                                                                                -2\\
                                                                              \end{array}
                                                                            \right),\qquad \varphi_4^{(1)}=\left(
                                                                              \begin{array}{c}
                                                                                1 \\
                                                                                1 \\
                                                                                0\\
                                                                                0\\
                                                                                1\\
                                                                              \end{array}
                                                                            \right),
$$
while $\varphi_2^{(1)}$ and $\varphi_5^{(1)}$ are very complicated and it is not worth giving here their explicit expression. Using our definitions we now find
$$h_2=\frac{3}{5}\,F^\dagger\,h_1\,F=\left(
                                                                                               \begin{array}{ccc}
                                                                                                 17/6 & 0 & \sqrt{5}/6 \\
                                                                                                 0 & 5 & 0 \\
                                                                                                 \sqrt{5}/6 & 0 & 13/6 \\
                                                                                               \end{array}
                                                                                             \right).
$$
Then $\sigma(h_2)=\{5,3,2\}$, which is again a proper subset of $\sigma(h_1)$. Moreover, if we compute $\varphi_j^{(2)}=F^\dagger\varphi_j^{(1)}$, we get
$$\varphi_1^{(2)}= \left(
                                                                              \begin{array}{c}
                                                                                0 \\
                                                                                5/3 \\
                                                                                0 \\
                                                                              \end{array}
                                                                            \right),\quad \varphi_3^{(2)}=-\frac{5}{2\sqrt{6}}\left(
                                                                              \begin{array}{c}
                                                                                \sqrt{5} \\
                                                                                0 \\
                                                                                1 \\
                                                                              \end{array}
                                                                            \right),\quad \varphi_4^{(2)}=\sqrt{\frac{5}{6}}\left(
                                                                              \begin{array}{c}
                                                                                -1 \\
                                                                                0 \\
                                                                                \sqrt{5}\\
                                                                              \end{array}
                                                                            \right).
$$
 We also see that $F^\dagger \varphi_2^{(1)}=F^\dagger \varphi_5^{(1)}=0$. This shows that, also in this example, the kernel of $F^\dagger$ contains non zero vectors and, as a consequence, that $FF^\dagger$ is not invertible, as it was already clear. Finally, it is just a simple computation to check that $h_2\varphi_1^{(2)}=5\varphi_1^{(2)}$, $h_2\varphi_3^{(2)}=3\varphi_3^{(2)}$, and $h_2\varphi_4^{(2)}=2\varphi_4^{(2)}$: the eigenvalues and their order are {\em respected}, as they must.

\subsection{A first infinite dimensional example}

Let now $\E=\{e_n,\,n\in\Bbb{N}\}$ be an orthonormal basis of the separable, infinite dimensional, Hilbert space $\Hil$. Let us introduce the following vectors: $\chi_{2n-1}:=\frac{1}{\sqrt{2}}e_n$, $\chi_{2n}:=\frac{1}{\sqrt{2}}e_n$, for all $n\geq 1$. Let $\F$ be the set of all these vectors. Hence $\F$ contains each vector of $\E$ twice, but the normalization of the various vectors is lost. It is easy to check that $\F$ is a Parceval tight frame: $\sum_{n\in\Bbb{N}}|<\chi_n,f>|^2=\|f\|^2$, for all $f\in\Hil$. We define the analysis and the synthesis operators as usual. Of course we have $F^\dagger F=\Id_\Hil$. As for $FF^\dagger$, it is easy to check that this operator can be represented as the following infinite matrix acting on $l^2(\Bbb{N})$:
$$
FF^\dagger=\left(
             \begin{array}{cccccccc}
               1 & 1 & 0 & 0 & 0 & 0 & .& . \\
               1 & 1 & 0 & 0 & 0 & 0 & .& . \\
               0 & 0 & 1 & 1 & 0 & 0 & .& . \\
               0 & 0 & 1 & 1 & 0 & 0 & .& . \\
               0 & 0 & 0 & 0 & 1 & 1 & .& . \\
               0 & 0 & 0 & 0 & 1 & 1 & .& . \\
               . & . & . & . & . & . & .& . \\
               . & . & . & . & . & . & .& . \\
             \end{array}
           \right)
$$
which can also be written as $FF^\dagger=\frac{1}{2}(\Id_{l^2}+P_2)$, where $P_2$ is the following permutation operator defined on $c\in l^2(\Bbb{N})$ as
$$
P_2c=P_2\left(
     \begin{array}{c}
       c_1 \\
       c_2 \\
       c_3 \\
       c_4 \\
       . \\
       . \\
     \end{array}
   \right)=\left(
     \begin{array}{c}
       c_2 \\
       c_1 \\
       c_4 \\
       c_3 \\
       . \\
       . \\
     \end{array}
   \right)
$$
It is clear that $P_2^{2n}=\Id_{l^2}$, and $P_2^{2n+1}=P_2$, for all $n\in\Bbb{N}$. It is also clear that $F^\dagger F$ is invertible but $FF^\dagger $ is not: the non zero (column) vector $d=(1,-1,0,0,0,0,\ldots)$ is mapped into the zero vector by $FF^\dagger $. Moreover $d$ belongs to $\ker(F^\dagger)$, as well as many others. Once again we cannot use Option I, but Option II still works. It is clear that the most general operator $\sum_{k=0}^\infty a_k P_2^k$ (notice that this might be just a formal series!) can be written as $\alpha\Id_{l^2}+\beta P_2$, and this is the expression for $h_1$ we are going to consider here. Indeed if $\alpha$ and $\beta$ are real, then $h_1=\alpha\Id_{l^2}+\beta P_2$ is self-adjoint and commutes with $\mathfrak{N}_1$. It is interesting to notice that $h_2$ in (\ref{26}) turns out to be a scalar operator. Indeed we find $h_2=\mathfrak{N}_2^{-1}\left(F^\dagger h_1 F\right)=F^\dagger \left(\alpha\Id_{l^2}+\beta P_2\right) F=\alpha F^\dagger F+\beta F^\dagger P_2 F$. But since $P_2=(2\mathfrak{N}_1-\Id_{l^2})$ and $F^\dagger F=\Id_\Hil$ we conclude that $h_2=(\alpha+\beta)\Id_\Hil$.

As for the relation between the eigenstates of $h_1$ and $h_2$, we just remark that the (column) vector $\varphi_1^{(1)}=(c_1,c_1,c_2,c_2,c_3,c_3,\ldots)$ is an eigenstate of $h_1$ for any non trivial choice of the $c_j$'s (some of these coefficients must be non zero). It satisfies the eigenvalue equation $h_1\varphi_1^{(1)}=(\alpha+\beta)\varphi_1^{(1)}$. But, because of the form of $h_2$, it is also clear that $\varphi_1^{(2)}=F^\dagger \varphi_1^{(1)}$ is eigenstate of $h_2$ with the same eigenvalue, $\epsilon_1=\alpha+\beta$.

\subsection{A second infinite dimensional example}

We use again the same set $\E$ as in Section III.3 to built up a different Parceval tight frame in the following way: we put, \cite{chri}, $\chi_1=e_1$, $\chi_2=\chi_3=\frac{1}{\sqrt{2}}e_2$, $\chi_4=\chi_5=\chi_6=\frac{1}{\sqrt{3}}e_3$, and so on. Then we call $\F=\{\chi_j,\,j\in\Bbb{N}\}$. We have  $\sum_{n\in\Bbb{N}}|<\chi_n,f>|^2=\|f\|^2$, for all $f\in\Hil$. We define the analysis and the synthesis operators as usual. Of course we have $F^\dagger F=\Id_\Hil$, while $FF^\dagger$ can be represented as the following infinite matrix acting on $l^2(\Bbb{N})$:
$$
FF^\dagger=\left(
             \begin{array}{cccccccc}
               1 & 0 & 0 & 0 & 0 & 0 & 0& . \\
               0 & 1/2 & 1/2 & 0 & 0 & 0 & 0& . \\
               0 & 1/2 & 1/2 & 0 & 0 & 0 & 0& . \\
               0 & 0 & 0 & 1/3 & 1/3 & 1/3 & 0& . \\
               0 & 0 & 0 & 1/3 & 1/3 & 1/3 & 0& . \\
               0 & 0 & 0 & 1/3 & 1/3 & 1/3 & 0& . \\
               0 & 0 & 0 & 0 & 0 & 0 & 1/4& . \\
               . & . & . & . & . & . & .& . \\
             \end{array}
           \right)
$$
It is evident that the non zero (column) vector $d=(0,1,-1,0,0,\ldots)$ is mapped into zero by $FF^\dagger$, which is therefore not invertible. It is also clear that $d$ belongs to $\ker(F^\dagger)$, as well as the  (column) vector $d=(0,0,0,1,-2,1,0,\ldots)$ and many others. Once again, as in all the examples considered so far, wa are forced to use Option II. This is linked to the fact that all the tight frames considered up to now are made of linearly dependent vectors. We put now $\mathfrak{N}_2=\Id_\Hil$ and $\mathfrak{N}_1=FF^\dagger$. It is easy to find examples of self-adjoint operators commuting with $\mathfrak{N}_1$. One such operator is the following diagonal infinite matrix: $h_1=diag(\alpha_1,\alpha_2,\alpha_2,\alpha_3,\alpha_3,\alpha_3,\ldots)$, with real $\alpha_j$'s. The orthonormal basis of eigenvectors of $h_1$ is the canonical basis in $l^2(\Bbb{N})$: $\hat\varphi_j^{(1)}=(0,0,\ldots,0,1,0,\ldots)$, where $1$ appears at the j-th place. Of course all the eigenvalues of $h_1$, but $\epsilon_1=\alpha_1$, are degenerate and the dimension of the related eigenspace increases with $n$: the degeneracy of $\epsilon_n$ is exactly $n$. As for $h_2$ and $\varphi_j^{(2)}$ we find, first of all, that $\varphi_j^{(2)}=F^\dagger \varphi_j^{(1)}=\chi_j$ which is therefore nothing but one of the vector of the original set $\E$ divided by some constant (depending on $j$). Since  $\mathfrak{N}_2=\Id_\Hil$, $h_2=F^\dagger h_1 F$, whose action on a generic $f\in\Hil$ can be written as
$$
h_2f=F^\dagger\left(h_1(Ff)\right)=\sum_{j\in\Bbb{N}}(h_1(Ff))_j\chi_j=\left(\sum_{j\in\Bbb{N}}\alpha_j|e_j><e_j|\right)f,
$$
where we have adopted the Dirac bra-ket notation. Then $h_2=\sum_{j\in\Bbb{N}}\alpha_j|e_j><e_j|$. Hence the eigenvalues of $h_1$ and $h_2$ coincide, with the difference that in $h_2$ they are all non degenerate! It is finally self-evident that $\varphi_j^{(2)}=\chi_j$ is an eigenstate of $h_2$.

A second choice of $h_1$, which also commutes with $\mathfrak{N}_1$, is given by the following operator
$$
h_1'=\left(
             \begin{array}{cccccccc}
               \alpha_1 & 0 & 0 & 0 & 0 & 0 & 0& . \\
               0 & \alpha_2 & \beta_2 & 0 & 0 & 0 & 0& . \\
               0 & \beta_2 & \alpha_2 & 0 & 0 & 0 & 0& . \\
               0 & 0 & 0 & \alpha_3 & \beta_3 & \beta_3 & 0& . \\
               0 & 0 & 0 & \beta_3 & \alpha_3 & \beta_3 & 0& . \\
               0 & 0 & 0 & \beta_3 & \beta_3 & \alpha_3 & 0& . \\
               0 & 0 & 0 & 0 & 0 & 0 & . & . \\
               . & . & . & . & . & . & .& . \\
             \end{array}
           \right)
$$
for all possible choices of real $\alpha_j$'s and $\beta_j$'s. The first eigenvectors of $h_1'$ are the following: $\varphi_1^{(1)}=(1,0,0,\ldots)$, $\varphi_2^{(1)}=(0,-1,1,0,0,\ldots)$, $\varphi_3^{(1)}=(0,1,1,0,0,\ldots)$,
$\varphi_4^{(1)}=(0,0,0,-\frac{1}{\sqrt{2}},0,\frac{1}{\sqrt{2}},\ldots)$, $\varphi_5^{(1)}=(0,0,0,-\frac{1}{\sqrt{6}},\sqrt{\frac{2}{3}},-\frac{1}{\sqrt{6}},\ldots)$,
$\varphi_6^{(1)}=\frac{1}{\sqrt{3}}(0,0,0,1,1,1,\ldots)$,
which corresponds respectively to the following eigenvalues: $\epsilon_1=\alpha_1$, $\epsilon_2=\alpha_2-\beta_2$, $\epsilon_3=\alpha_2+\beta_2$, $\epsilon_4=\alpha_3-\beta_3$, $\epsilon_5=\alpha_3-\beta_3$, $\epsilon_6=\alpha_3+2\beta_3$, and so on. The related eigenstates of $h_2$ turn out to be
$\varphi_1^{(2)}=F^\dagger \varphi_1^{(1)}=e_1$,  $\varphi_3^{(2)}=F^\dagger \varphi_3^{(1)}=\sqrt{2}\,e_2$, $\varphi_6^{(2)}=F^\dagger \varphi_6^{(1)}=e_3$, while $\varphi_2^{(2)}=\varphi_4^{(2)}=\varphi_5^{(2)}=0$. The computation of $h_2'=F^\dagger h_1' F$ produces now $h_2'=\sum_{j\in\Bbb{N}}\tilde\alpha_j|e_j><e_j|$, where $\tilde\alpha_1=\alpha_1$, $\tilde\alpha_2=\alpha_2+\beta_2$, $\tilde\alpha_3=\alpha_3+2\beta_3$, $\tilde\alpha_4=\alpha_4+3\beta_4$ and so on. We see that the $e_j$'s are eigenstates of $h_2'$ but with different eigenvalues with respect to those of $h_2$.  This is in agreement with the fact that $h_1'$ and $h_2'$ must be almost isospectral.

\subsection{What if we start with an orthonormal basis?}

In all the previous examples we have been forced to use Option II because $FF^\dagger$ turned out to be not invertible. As we have already remarked, this is related to the fact that all the frames considered above consist of linearly dependent vectors. Here we consider the case of a tight frame of linearly independent vectors. The most natural choice of such a set is simply an orthonormal basis. So we take $\F=\E=\{e_j,\,j\in\Bbb{N}\}$, and we construct the standard analysis and synthesis operators $F:\Hil\rightarrow l^2(\Bbb{N})$ and $F^\dagger: l^2(\Bbb{N})\rightarrow\Hil$. Then we have
$$
F^\dagger F=\Id_\Hil,\qquad FF^\dagger=\Id_{l^2}
$$
It follows that both $F^\dagger F$ and $FF^\dagger$ are trivially invertible, so that both Option I and Option II are available and will be considered now.

\vspace{2mm}

{\bf Option I:} In this case we put $X=F^\dagger$, $N_1=XX^\dagger=F^\dagger F=\Id_\Hil$, $N_2=X^\dagger X=FF^\dagger =\Id_{l^2}$. Moreover $[h_1,N_1]=0$ for all possible choices of $h_1$. It is clear that $h_2=Fh_1F^\dagger$ is an operator on $l^2(\Bbb{N})$. It is interesting to consider the situation in which $e_j$ are exactly the eigenstates of $h_1$: $e_j\equiv\varphi_j^{(1)}$, $h_1\varphi_j^{(1)}=\epsilon_j\varphi_j^{(1)}$, $j\in\Bbb{N}$. Let now $c\in l^2(\Bbb{N})$. We have
$$
h_2c=Fh_1F^\dagger c=\left\{<\varphi_j^{(1)},h_1F^\dagger c>\right\}_{j\in\Bbb{N}}=
\left\{<h_1\varphi_j^{(1)},F^\dagger c>\right\}_{j\in\Bbb{N}}=$$
$$=\left\{\epsilon_j<\varphi_j^{(1)},F^\dagger c>\right\}_{j\in\Bbb{N}}=
\left\{\epsilon_j<\varphi_j^{(1)},\sum_{k\in\Bbb{N}}c_k\varphi_k^{(1)}>\right\}_{j\in\Bbb{N}}=\left\{\epsilon_j\,c_j
\right\}_{j\in\Bbb{N}}
$$
It could happen that $\sum_{n\in\Bbb{N}}|\epsilon_n\,c_n|^2=\infty$. If this happens, then $h_2$ is unbounded. Otherwise $h_2$ is a bounded operator from $l^2(\Bbb{N})$ in itself.

As for the eigenstates of $h_2$ we find that
$$
\varphi_k^{(2)}=X^\dagger \varphi_k^{(1)}=F\varphi_k^{(1)}=\left\{<\varphi_j^{(1)},\varphi_k^{(1)}>\right\}_{j\in\Bbb{N}}=
\left\{\delta_{j,k}\right\}_{j\in\Bbb{N}}=(0,0,\ldots,0,1,0,0,\ldots),
$$
where 1 is in the k-th place. This result is in agreement with our general scheme. Indeed, using the above result for $h_2c$, we find that $h_2\varphi_k^{(2)}=(0,0,\ldots,0,\epsilon_k,0,0,\ldots)=\epsilon_k\varphi_k^{(2)}$. It is interesting to observe also that since $F$ maps an orthonormal basis of $\Hil$ into an orthonormal basis of $l^2(\Bbb{N})$, $F$ is necessarily an  unitary operator.

\vspace{2mm}

{\bf Option II:} Now we put $X=F:\Hil\rightarrow l^2(\Bbb{N})$, $\mathfrak{N}_1=XX^\dagger=FF^\dagger =\Id_{l^2}$, $\mathfrak{N}_2=X^\dagger X=F^\dagger F=\Id_{\Hil}$. Moreover $[h_1,\mathfrak{N}_1]=0$ for all possible choices of $h_1$. We should stress that, while in Option I $h_1$ is an operator on $\Hil$, here is an operator on $l^2(\Bbb{N})$: the role of $\Hil$ and $l^2(\Bbb{N})$ are exchanged, and
the hamiltonian $h_2=F^\dagger h_1 F$ {\em lives} in $\Hil$. Following the same idea as in Option I we take as vectors in $\F$ the eigenvectors of $h_2$: $e_j\equiv\varphi_j^{(2)}$, $h_2\varphi_j^{(2)}=\epsilon_j\varphi_j^{(2)}$, $j\in\Bbb{N}$. In this case it is convenient to consider $h_2$ as the starting point, and try to recover $h_1$ from $h_2$. This is possible since, in our conditions, $h_2=F^\dagger h_1 F$ immediately implies that $Fh_2F^\dagger=FF^\dagger h_1 FF^\dagger=\Id_{l^2}h_1\Id_{l^2}=h_1$.
Hence we have
$$
h_1\left(F\varphi_j^{(2)}\right)=Fh_2F^\dagger F\varphi_j^{(2)}=Fh_2\Id_{l^2}\varphi_j^{(2)}=\epsilon_j\left(F\varphi_j^{(2)}\right),
$$
which shows that $F\varphi_j^{(2)}$ is an eigenstate of $h_1$ with eigenvalue $\epsilon_j$, as expected for the general reasons discussed in Section I. These vectors of $l^2(\Bbb{N})$ can be written explicitly recalling the definition of the operator $F$. We find that
$$
\varphi_k^{(1)}=F\varphi_k^{(2)}=\left\{<\varphi_j^{(2)},\varphi_k^{(2)}>\right\}_{j\in\Bbb{N}}=
\left\{\delta_{j,k}\right\}_{j\in\Bbb{N}}=(0,0,\ldots,0,1,0,0,\ldots),
$$
where 1 is in the k-th place. Once more, as in Option I, the image of the starting orthonormal basis of $\Hil$ is the canonical basis of $l^2(\Bbb{N})$. Again, this implies that $F$ is unitary.

\vspace{3mm}
We close this section remarking that the unitarity of $F$ implies that the example we are considering here is just a formal version of that given at the end of Section II and contained in \cite{alibag}.

\section{Examples from tight g-frames}

Recently the notion of frames has been extended in order to unify the existent literature on generalized frames, \cite{sun,najati}. This section is devoted to the analysis of some examples of our procedure arising from the  theory of g-frames, which we now quickly review in a slightly simplified form for reader's convenience.

Let $\Hil$ and $\tilde\Hil$ be two (in general) different Hilbert spaces, $\J$ a discrete set of indexes and let $\Lc=\{\Lambda_j:\Hil\rightarrow\tilde\Hil,\,j\in\J\}$ be a set of operators mapping $\Hil$ into $\tilde\Hil$. $\Lc$ is called {\em an $(A,B)$ g-frame of $(\Hil,\tilde\Hil)$} if there exist two positive numbers $A$ and $B$, with $0<A\leq B<\infty$, such that for all $f\in\Hil$
\be
 A\|f\|_\Hil^2 \leq \sum_{j\in\J}\|\Lambda_jf\|_{\tilde\Hil}^2 \leq B\|f\|_\Hil^2
\label{41}
\en
In particular a g-frame is called {\em tight} if $A=B$ and it is called a {\em Parceval g-frame} if $A=B=1$. Standard frames are recovered if we have $\tilde\Hil=\Bbb{C}$ and $\Lambda_j=<\varphi_j,.>$, with $\varphi_j$ belonging to a given $(A,B)$-frame of $\Hil$.
Let us now define a new Hilbert space, which looks like an  $l^2(\Bbb{N})$ space where the sequences of complex numbers
are replaced by sequences of elements of $\tilde\Hil$:
\be
\hat\Hil:=\left\{\underline{f}:=\{f_j\in\tilde\Hil\}_{j\in\J}, \mbox{ such that }\|\underline{f}\|_{\hat\Hil}^2:=
\sum_{j\in\J}\|f_j\|_{\tilde\Hil}^2<\infty\right\}
\label{42}\en
with the following scalar product:
\be
<\underline{f},\underline{g}>_{\hat\Hil}:=\sum_{j\in\J}<f_j,g_j>_{\tilde\Hil}.
\label{43}\en
Now we can associate to the set $\Lc$  a bounded operator $F_g:\Hil \rightarrow \hat\Hil$, called again the {\em analysis operator},  defined by the formula
\be
\forall f\in \Hil \hspace{1cm}  (F_gf)_j = \Lambda_j\, f, \qquad\Longrightarrow\qquad F_gf=\{\Lambda_j\, f\}_{j\in\J}.
\label{44}
\en
The vector $(F_gf)_j$ belongs to $\tilde\Hil$ for each $j\in\J$, while $F_gf$ belongs to $\hat\Hil$. As for standard frames, we find that $\|F_g\| \leq \sqrt{B}$. The
adjoint of the operator $F_g$, the so-called {\em synthesis operator} $F_g^\dagger$,  maps $\hat\Hil$ into $\Hil$, and is such that
 \be
   F_g^\dagger \underline{f} = \sum_{j\in\J}\Lambda_j^\dagger \,f_j
\label{45}
\en
Using these two operators we can define the g-frame operator $S_g=F_g^\dagger\,F_g$ which acts on a generic element $f\in\Hil$ as
\be
S_gf=F_g^\dagger\,F_gf=\sum_{j\in\J}\Lambda_j^\dagger\,\Lambda_j\,f
\label{46}\en

By means of these operators condition (\ref{41}) can be rewritten in the following
equivalent way: $\Lc$ is an (A,B) {g-frame} of $(\Hil,\tilde\Hil)$ if there exist
two positive constants $A$ and $B$, $0<A\leq B<\infty$, such that the inequalities
\be
A\Id_\Hil \leq S_g \leq B \Id_\Hil
\label{47}
\en
hold in the sense of the operators.  Repeating the same procedure as in the previous section we deduce that $S_g$ is a self-adjoint, positive and bounded operator. More than this, we find that because of (\ref{47}), $A\leq\|S_g\|\leq B$ holds. $S^{-1}$ clearly exists in $\Hil$, and we find that $B^{-1}\Id_\Hil \leq S_g^{-1} \leq A^{-1} \Id_\Hil$. $S_g^{-1}$ and $S_g$ can be used together now to get two resolutions of the identity of $\Hil$ and their related
expansions of any given vector of $\Hil$. Indeed, defining a new operator $\tilde\Lambda_j:=\Lambda_jS_g^{-1}$ mapping $\Hil$ into $\tilde\Hil$, and its adjoint $\tilde\Lambda_j^\dagger:=S_g^{-1}\Lambda_j^\dagger:\tilde\Hil\rightarrow\Hil$, we find that
\be
f=S_gS_g^{-1}f=\sum_{j\in\J}\Lambda_j^\dagger\,\tilde\Lambda_j \,f
\quad\mbox{ or }\quad f=S_g^{-1}S_gf=\sum_{j\in\J}\tilde\Lambda_j^\dagger\,\Lambda_j \,f,
\label{48}\en
which are the extended version of (\ref{36}). The {\em dual set} of $\Lc$, $\tilde\Lc=\{\tilde\Lambda_j,\,j\in\J\}$, is a g-frame by itself, and in particular is a $\left(\frac{1}{B},\frac{1}{A}\right)$ g-frame, whose dual coincides with $\Lc$ itself. As in the previous section, our interest here is mainly in the tight g-frames, since in this case $S_g$ is proportional to the identity operator.

\subsection{Example number one}

Let $\Hil=\tilde\Hil=\Lc^2(\Bbb{R})$, and let, calling $I_j=[j,j+1[$, $j\in\Bbb{Z}$,
$$
\left(\Lambda_jf\right)(x)=\left\{
\begin{array}{ll}
f(x), \qquad x\in I_j  \\
0,\hspace{14mm} \mbox{otherwise}\\
\end{array}
\right\}=\chi_j(x)f(x)
$$
for all $f(x)\in\Lc^2(\Bbb{R})$. Here $\chi_j(x)$ is the characteristic function in $I_j$. It is clear that $\Lambda_j:\Lc^2(\Bbb{R})\rightarrow\Lc^2(\Bbb{R})$ and it is also easy to check that $\Lc=\{\Lambda_j,\,j\in\Bbb{Z}\}$ is a Parceval g-frame of $(\Lc^2(\Bbb{R}),\Lc^2(\Bbb{R}))$. Indeed we find that, for all $f(x)\in\Lc^2(\Bbb{R})$, $\sum_{j\in\Bbb{Z}}\|\Lambda_jf\|_{\tilde\Hil}^2=\|f\|_\Hil^2$. Moreover, the adjoint of $\Lambda_j$ coincides with $\Lambda_j$ and, since $S_g=\Id_\Hil$, the dual g-frame of $\Lc$ coincides with $\Lc$ itself. Defining now $\hat\Hil$, $F_g$ and $F_g^\dagger$ as in (\ref{42}), (\ref{44}) and (\ref{45}), we can check that $F_g^\dagger\, F_g=\Id_\Hil$ while $F_g\,F_g^\dagger$ acts on $\hat\Hil$ as the following infinite matrix: $F_g\,F_g^\dagger=diag\{\ldots,\chi_{-1}(x),\chi_0(x),\chi_1(x),\ldots\}$, which behaves as a projector operator on $\hat\Hil$. Indeed we have, in particular, $\left(F_g\,F_g^\dagger\right)^2=F_g\,F_g^\dagger$. We now take $X=F_g$, $\mathfrak{N}_2=
F_g^\dagger\,F_g=\Id_\Hil$ and $\mathfrak{N}_1=F_g\,F_g^\dagger$. It is clear that the  operator on $\hat\Hil$ defined as $diag\{\ldots,\mu_{-1}(x),\mu_0(x),\mu_1(x),\ldots\}$ commutes with $\mathfrak{N}_1$ for all choices of the real functions $\mu_j(x)$. Hence we can take this infinite diagonal matrix as our hamiltonian $h_1$: $[h_1,\mathfrak{N}_1]=0$. From (\ref{26}) with $X=F_g$ we deduce that $h_2=F_g^\dagger h_1 F_g=\sum_{j\in\Bbb{Z}}\mu_j(x)\,\chi_j(x)$, which is an operator on $\Lc^2(\Bbb{R})$.

As for the relation between the eigenvectors of $h_1$ and $h_2$, let us assume for simplicity that $\mu_0(x)=\mu_0$. Hence the (column) vector of $\hat\Hil$ $\varphi_0^{(1)}=(\ldots,0,0,\chi_0(x),0,0,\ldots)$ satisfies the eigenvalue equation $h_1\varphi_0^{(1)}=\mu_0\varphi_0^{(1)}$. Therefore $\epsilon_0=\mu_0$. Let us now define, as in (\ref{26}), $\varphi_0^{(2)}=F_g^\dagger \varphi_0^{(1)}=\Lambda_0^\dagger\chi_0(x)=\left(\chi_0(x)\right)^2=\chi_0(x)$. It is trivial to check that $h_2\varphi_0^{(2)}=\mu_0\varphi_0^{(2)}$, as expected from our general results.

\subsection{Example number two}

In the previous example the characteristic functions $\chi_j(x)$ behave as projectors acting on $\Lc^2(\Bbb{R})$. This suggests the following extension: let $\Hil=\tilde\Hil$ be a given Hilbert space, and $\{P_j,\,j\in\Bbb{N}\}$ a family of orthogonal projections satisfying $P_jP_k=\delta_{j,k}P_j$, $P_j^\dagger=P_j$ and such that $\sum_{j\in\Bbb{N}}P_j=\Id_\Hil$. If we now define $\Lambda_j=P_j$ for all $j\in\Bbb{N}$, we can check that $\Lc=\{\Lambda_j,\,j\in\Bbb{N}\}$ is a Parceval g-frame of $(\Hil,\Hil)$: $\sum_{j\in\Bbb{N}}\|\Lambda_j f\|_{\tilde\Hil}^2=\|f\|_\Hil^2$ for all $f\in\Hil$. Notice that we are still using $\|.\|_{\tilde\Hil}$ even if it coincides with $\|.\|_{\Hil}$, simply to keep in mind the possible differences between the two norms. As in the previous example the g-frame operator $S_g$ is simply $\Id_\Hil$ and therefore the dual g-frame coincides with $\Lc$. We now introduce $\hat\Hil$ and the synthesis and analysis operators as usual:
$$
F_g:\Hil\rightarrow\hat\Hil, \qquad F_gf=\{P_jf\}_{j\in\Bbb{N}},
$$
$$
F_g^\dagger:\hat\Hil\rightarrow\Hil, \qquad F_g^\dagger \underline{f}=\sum_{j\in\Bbb{N}}P_jf_j,
$$
and $S_g=F_g^\dagger\,F_g=\Id_\Hil$ while $F_gF_g^\dagger \underline{f}=\{P_jf_j\}_{j\in\Bbb{N}}$. This implies that $F_gF_g^\dagger$ can be represented as the following infinite dimensional matrix: $F_gF_g^\dagger=diag(P_1,P_2,P_3,\ldots)$, which is not invertible in $\hat\Hil$.

Let us now take $X=F_g$. Hence we get  $\mathfrak{N}_2=
F_g^\dagger\,F_g=\Id_\Hil$ and $\mathfrak{N}_1=F_g\,F_g^\dagger=diag(P_1,P_2,P_3,\ldots)$, which commutes with the operator $h_1=\left(\sum_{j\in\Bbb{N}}\alpha_j^{(1)}P_j,\sum_{j\in\Bbb{N}}\alpha_j^{(2)}P_j,\ldots\right)$ for all possible choices of the real coefficients $\alpha_j^{(k)}$. The operator $h_2=F_g^\dagger h_1 F_g$ turns our to be $h_2=\sum_{j\in\Bbb{N}}\alpha_j^{(j)}P_j$, which acts on $\Hil$.

The analysis of the eigenvectors of $h_1$ and $h_2$ goes like this:

let us assume first that, to make life easier, $\alpha_j^{(1)}=0$ for all $j\neq 2$ and that  $\alpha_2^{(1)}\neq0$. In this case the (column) vector $\varphi_0^{(1)}=(P_2f,0,0,\ldots)$ is an eigenstate of $h_1$ for all choices of $f\in\Hil$ such that $f\notin \ker(P_2)$: $h_1\varphi_0^{(1)}=\alpha_2^{(1)}\varphi_0^{(1)}$. However, $\varphi_0^{(2)}=F_g^\dagger \varphi_0^{(1)}=P_1\left(P_2f\right)=0$.

This suggests to consider a different situation, the one in which $\alpha_j^{(1)}=0$ for all $j\neq 1$ and that  $\alpha_1^{(1)}\neq0$. Now $\varphi_0^{(1)}=(P_1f,0,0,\ldots)$ is an eigenstate of $h_1$ with eigenvalue $\alpha_1^{(1)}$ for all choices of $f\in\Hil$ such that $f\notin \ker(P_1)$. Moreover $\varphi_0^{(2)}=F_g^\dagger \varphi_0^{(1)}=P_1\left(P_1f\right)=P_1 f$, and it is clear that $h_2\varphi_0^{(2)}=\alpha_1^{(1)}\varphi_0^{(2)}$, as expected.  Generalizing this example is trivial and will not be done here.

\subsection{Example number three}

Let $\{P_j,\,j\in\Bbb{N}\}$ be as above and $V$ a map between two (in general) different Hilbert spaces $\Hil$ and $\tilde\Hil$ such that a real numbers $A>0$  exists for which $V^\dagger\,V=A\Id_{\Hil}$. Then, putting $\Lambda_j=VP_j$, the set $\Lc=\{\Lambda_j,\,j\in\Bbb{N}\}$ is a tight g-frame of $(\Hil,\tilde\Hil)$: $\sum_{j\in\Bbb{N}}\|\Lambda_j f\|_{\tilde\Hil}^2=A\|f\|_\Hil^2$ for all $f\in\Hil$. Introducing $\hat\Hil$ as usual, the synthesis and analysis operators look like
$$
F_g:\Hil\rightarrow\hat\Hil, \qquad F_gf=\{VP_jf\}_{j\in\Bbb{N}},
$$
$$
F_g^\dagger:\hat\Hil\rightarrow\Hil, \qquad F_g^\dagger \underline{f}=\sum_{j\in\Bbb{N}}P_jV^\dagger f_j,
$$
and we find that $S_g=F_g^\dagger\,F_g=A\Id_\Hil$ while $F_gF_g^\dagger \underline{f}=\{VP_jV^\dagger f_j\}_{j\in\Bbb{N}}$. This implies that $F_gF_g^\dagger$ can be represented as the following infinite dimensional matrix: $F_gF_g^\dagger=diag(VP_1V^\dagger,VP_2V^\dagger,VP_3V^\dagger,\ldots)$, which is not invertible in $\hat\Hil$.

As a concrete example of operator $V$ we can consider here $V=F$, where $F$ is the operator introduced in Section III.3. Hence we have $\mathfrak{N}_1=F_gF_g^\dagger=diag(FP_1F^\dagger,FP_2F^\dagger,FP_3F^\dagger,\ldots)$. Using the results of the previous example is clear that $\mathfrak{N}_1$ commutes with $$h_1=\left(F\left(\sum_{j\in\Bbb{N}}\alpha_j^{(1)}P_j\right)F^\dagger,
F\left(\sum_{j\in\Bbb{N}}\alpha_j^{(2)}P_j\right)F^\dagger,\ldots\right),$$ for all possible choices of the real coefficients $\alpha_j^{(k)}$. The operator $h_2=F_g^\dagger h_1 F_g$ turns our to be $h_2=\sum_{j\in\Bbb{N}}\alpha_j^{(j)}P_j$, which coincides with the operator found in the previous example. Moreover, using the results obtained above, it is easy to verify the following relation between the eigenstates of $h_1$ and $h_2$: let us suppose that $\alpha_j^{(1)}=0$ for all $j\neq 1$ and that  $\alpha_1^{(1)}\neq0$. Now $\varphi_0^{(1)}=(FP_1f,0,0,\ldots)$ is an eigenstate of $h_1$ with eigenvalue $\alpha_1^{(1)}$ for all choices of $f\in\Hil$ such that $f\notin \ker(P_1)$. Moreover $\varphi_0^{(2)}=F_g^\dagger \varphi_0^{(1)}=P_1F^\dagger\left(FP_1f\right)=P_1 f$, and it is clear that $h_2\varphi_0^{(2)}=\alpha_1^{(1)}\varphi_0^{(2)}$, as expected. Again, an analogous analysis could be carried out for different choices of the $\alpha_j^{(k)}$'s, but this will not be done here.

\section{Conclusions}

We have discussed a general procedure which generates, starting from a set of {\em minimal} igreedients, a pair of self-adjoint operators living in different Hilbert spaces which are almost isospectral and whose eigenvectors are related by an intertwining operator. We have applied this procedure to many examples arising from frames and g-frames, which naturally provide examples of intertwining operators between different Hilbert spaces.

These are essentially mathematical applications. Our next step will consist in looking for more physically relevant applications, other than the one in \cite{alibag}, and for the extension of our procedure to non-isospectral hamiltonians, in the direction already suggested in some papers, see \cite{spi,fern} for instance.

As we have already mentioned before, physical applications of a certain interest are surely those involving differential operators  in connection with some Schr\"odinger equation. We have chosen not to consider these examples here since they are more naturally related to a single Hilbert space and since, in any case, they deserve a review just for themselves. This is part of our future projects.

\section*{Acknowledgements}

  The author acknowledges financial support by the Murst, within the  project {\em Problemi
Matematici Non Lineari di Propagazione e Stabilit\`a nei Modelli
del Continuo}, coordinated by Prof. T. Ruggeri.

\end{document}